\documentclass[a4paper,11pt]{article}
\usepackage{pos}
\usepackage{subfig}
\usepackage{orcidlink}
\usepackage{hyperref}
\usepackage{amsmath}
\usepackage{adjustbox}
\usepackage{verbatim}
\graphicspath{{./plots/},{./plots/mixing_angle/},{./plots/spectrum/},{./plots/effective_mass/}}

\newcommand{\orcidauthorBENNETT}{0000-0002-1678-6701}
\newcommand{\orcidauthorFORZANO}{0000-0003-0985-8858}
\newcommand{\orcidauthorHONG}{0000-0002-3923-4184}
\newcommand{\orcidauthorHSIAO}{0000-0002-8522-5190}
\newcommand{\orcidauthorLEE}{0000-0002-4616-2422}
\newcommand{\orcidauthorLIN}{0000-0003-3743-0840}
\newcommand{\orcidauthorLUCINI}{0000-0001-8974-8266}
\newcommand{\orcidauthorPIAI}{0000-0002-2251-0111} 
\newcommand{\orcidauthorVADACCHINO}{0000-0002-5783-5602}
\newcommand{\orcidauthorZIERLER}{0000-0002-8670-4054}

\title{Progress on pseudoscalar flavour-singlets in Sp(4) with mixed fermion representations}

\author*[a]{Fabian Zierler\,\orcidlink{\orcidauthorZIERLER}}
\emailAdd{fabian.zierler@swansea.ac.uk}
\author[b]{Ed Bennett\,\orcidlink{\orcidauthorBENNETT}}
\author[b]{Niccolò Forzano\,\orcidlink{\orcidauthorFORZANO}}
\author[c,d]{Deog~Ki Hong\,\orcidlink{\orcidauthorHONG}}
\author[e]{Ho Hsiao\,\orcidlink{\orcidauthorHSIAO}}
\author[f]{Jong-Wan Lee\,\orcidlink{\orcidauthorLEE}}
\author[e,h]{C.-J. David Lin\,\orcidlink{\orcidauthorLIN}}
\author[b,i]{Biagio~Lucini\,\orcidlink{\orcidauthorLUCINI}}
\author[a]{Maurizio Piai\,\orcidlink{\orcidauthorPIAI}}
\author[j]{Davide Vadacchino\,\orcidlink{\orcidauthorVADACCHINO}}

\affiliation[a]{Department of Physics, Faculty  of Science and Engineering, Swansea University, Singleton Park, SA2 8PP, Swansea, Wales, United Kingdom}
\affiliation[b]{Swansea Academy of Advanced Computing, Swansea University (Bay Campus), Fabian Way, SA1 8EN Swansea, Wales, United Kingdom}
\affiliation[c]{Department of Physics, Pusan National University, Busan 46241, Korea}
\affiliation[d]{Extreme Physics Institute, Pusan National University, Busan 46241, Korea}
\affiliation[e]{Institute of Physics, National Yang Ming Chiao Tung University, 1001 Ta-Hsueh Road, Hsinchu 30010, Taiwan}
\affiliation[f]{Particle Theory  and Cosmology Group, Center for Theoretical Physics of the Universe, Institute for Basic Science (IBS), Daejeon, 34126, Korea }
\affiliation[h]{Centre for High Energy Physics, Chung-Yuan Christian University, Chung-Li 32023, Taiwan}
\affiliation[i]{Department of Mathematics, Faculty of Science and Engineering, Swansea University (Bay Campus), Fabian Way, SA1 8EN Swansea, Wales, United Kingdom}
\affiliation[j]{Centre for Mathematical Sciences, University of Plymouth, Plymouth, PL4 8AA, United Kingdom}

\abstract{We measure the masses of the pseudoscalar flavour-singlet meson states in the $Sp(4)$ gauge theory coupled to two Dirac fermions transforming in the fundamental representation and three Dirac fermions in the antisymmetric representation. This theory provides a compelling ultraviolet completion for the minimal composite  Higgs model implementing also partial compositeness  for the top quark. The spectrum contains two, comparatively light,  pseudoscalar flavour-singlet states, which mix with one another. One of them is a Nambu-Goldstone boson (in the massless limit), whereas the other  receives a mass from the $U(1)_A$ axial anomaly. We demonstrate how to measure the mixing between these two states. For moderately heavy fermion masses, we find that  the two wave functions are dominated by one of the fermion representations,  mixing effects being small.}

\FullConference{The 41st International Symposium on Lattice Field Theory (LATTICE2024)\\
 28 July - 3 August 2024\\
Liverpool, UK\\}

\begin{document}
\maketitle

\section{Introduction}

Gauge theories with matter field content  transforming as an admixture of
representations
of the gauge group have been proposed as short-distance completions of
composite Higgs models that implement also partial compositeness for the top quark~\cite{Kaplan:1983fs,Georgi:1984af, Dugan:1984hq, Kaplan:1991dc}. In these models, the Higgs boson is a composite state of fermions (hyperquarks) charged under a new, non-Abelian gauge symmetry mediated by additional gauge bosons (hypergluons). 
The global symmetries of these theories factorize into two (non-Abelian) cosets, one of which is related to the physics of the Higgs boson, and the other to the properties of the heavy quarks (the top).\footnote{See Ref.~\cite{Vecchi:2015fma} for a model in which this can be achieved with one species of fermions.}
The $Sp(4)$ gauge theory coupled to $N_{\rm as}=3$  Dirac fermions transforming in the two-index antisymmetric representation  and $N_{\rm f}=2$ in the fundamental is the minimal such theory 
that meets all phenomenological requirements, and can be tested numerically on the lattice~\cite{Barnard:2013zea}---see also 
the reviews~\cite{Panico:2015jxa,Witzel:2019jbe,Cacciapaglia:2020kgq}, and the summary tables in Refs.~\cite{Ferretti:2013kya,Ferretti:2016upr,Cacciapaglia:2019bqz}.

Interest in new gauge theories with $Sp(2N)$ gauge group has motivated the development of the research programme of \textit{Theoretical Explorations on the Lattice with Orthogonal and Symplectic groups} (TELOS)~\cite{Bennett:2017kga,Bennett:2019jzz,Bennett:2019cxd,Bennett:2020hqd,Bennett:2020qtj,Bennett:2022yfa,Bennett:2022gdz,Bennett:2022ftz, Bennett:2023wjw,Bennett:2023gbe,Bennett:2023mhh, Bennett:2023qwx,Bennett:2024cqv,Bennett:2024wda}---see also Refs.~\cite{Kulkarni:2022bvh,Bennett:2023rsl,Dengler:2024maq,Bennett:2024bhy} 
for related applications to dark matter models---and the work reported in this contribution, and the associated publication~\cite{Bennett:2024wda}, are part of this programme. Other lattice studies of gauge theories with mixed representations include the $SU(4)$ gauge theory with fundamental and antisymmetric Dirac fermions~\cite{DeGrand:2016mxr,Ayyar:2017qdf,Ayyar:2017uqh,Ayyar:2017vsu,Ayyar:2018ppa,Ayyar:2018zuk,Cossu:2019hse,Lupo:2021nzv,DelDebbio:2022qgu,Golterman:2020pyx,Hasenfratz:2023sqa}, and $SU(2)$ gauge theory with fundamental and adjoint matter~\cite{Bergner:2020mwl, Bergner:2021ivi}.

A distinctive feature of gauge theories with matter in an admixture of multiple representations is the occurrence of an additional, flavour-singlet, Abelian pseudo-Nambu-Goldstone boson (pNGB), with peculiar phenomenological properties~\cite{Belyaev:2015hgo, DeGrand:2016pgq}. This is due to the existence of an independent axial $U(1)$ current for each fermion representation. Only one combination of  those currents is broken by the axial anomaly. The associated state (which in QCD can be identified with the $\eta'$ meson) acquires a mass even in the limit in which the hyperquarks are massless~\cite{Witten:1979vv,Witten:1978bc,Veneziano:1979ec}, while the remaining flavour-singlet pseudoscalar state does not. This phenomenology of the additional pNGB has been studied in the context of composite Higgs models~\cite{BuarqueFranzosi:2021kky,Cacciapaglia:2019bqz,Belyaev:2016ftv}. The relevant leading-order description within chiral perturbation theory has been developed in Ref.~\cite{DeGrand:2016pgq}. For the purposes of this work we refer to the light pNGB state as $\eta'_{l}$, while the state associated with the axial anomaly is denoted as $\eta'_{h}$.
This contribution reports the highlights of  non-perturbative lattice investigations into the masses of the $\eta'_{l}$ and $\eta'_{h}$, as well as their mixing angle, $\phi$, and we refer the reader to Ref.~\cite{Bennett:2024wda} for details and extensive discussions. 

\section{Lattice field theory observables}

The (Minkowski space) Lagrangian  density of the strongly interacting theory is given by
\begin{align}
    \mathcal L = -\frac{1}{2} {\rm Tr}\, G_{\mu\nu} G^{\mu\nu} +  \sum_{I=1}^{2} \bar Q^I \left( i\gamma^\mu D_\mu - m^{\rm f} \right) Q^I+ \sum_{K=1}^{3} \bar \Psi^K \left( i\gamma^\mu D_\mu - m^{\rm as} \right) \Psi^K,
\end{align}
where $Q$ denotes the fermions transforming in the fundamental representation and $\Psi$ those in the antisymmetric one. We discretize the theory as described in Sect.~\ref{Sec:num}.
We perform lattice measurements using gauge-invariant interpolating operators built with two fermions, transforming in a single representation. We define
\begin{align} \label{eq:ps_operators}
    O_{\eta^{\rm as}}(x) &\equiv \frac{1}{\sqrt{N_{\rm as}}} \sum_{K=1}^{N_{\rm as}} \bar \Psi^K(x) \gamma_5 \Psi^K(x), &
    O_{\eta^{\rm f}}(x)  &\equiv \frac{1}{\sqrt{N_{\rm f}}} \sum_{I=1}^{N_{\rm f}} \bar  Q^I(x)   \gamma_5  Q^I(x)\,.
\end{align}
From these operators we build a correlation matrix for the pseudoscalar flavour-singlet sector. Diagrammatically, the correlation matrix has the following form
\begin{align} \label{eq:correlation_matrix}
    C(x,y) = 
    \begin{pmatrix} 
    \langle \bar O_{\eta^{\rm as}} O_{\eta^{\rm as}} \rangle & \!\!\! \langle \bar O_{\eta^{\rm f}} O_{\eta^{\rm as}} \rangle \\ 
    \langle \bar O_{\eta^{\rm as}} O_{\eta^{\rm f}}  \rangle & \!\!\! \langle \bar O_{\eta^{\rm  f}} O_{\eta^{\rm  f}} \rangle
    \end{pmatrix}
    = 
    \begin{pmatrix}
    - \vcenter{\hbox{\includegraphics[scale=0.35,page=31]{contractions.pdf}}} + {N_{\rm as}}
    ~ \vcenter{\hbox{\includegraphics[scale=0.35,page=33]{contractions.pdf}}} & \!\!\!\!\!\!
    {\sqrt{N_{\rm as} N_{\rm f}} } \vcenter{\hbox{\includegraphics[scale=0.37,page=32]{contractions.pdf}}} \\
    {\sqrt{N_{\rm as} N_{\rm f}} } \vcenter{\hbox{\includegraphics[scale=0.37,page=32]{contractions.pdf}}} & \!\!\!\!\!\!
    - \vcenter{\hbox{\includegraphics[scale=0.35,page=30]{contractions.pdf}}} + {N_{\rm f}}   
    ~ \vcenter{\hbox{\includegraphics[scale=0.35,page=34]{contractions.pdf}}}
    \end{pmatrix}\,.
\end{align}

We further enlarge the correlation matrix $C(x,y)$ by including several layers of Wuppertal smearing on the operators, as described in Ref.~\cite{Bennett:2024cqv}. We extract the energy levels  by performing a variational analysis, within the context of a generalized eigenvalue problem (GEVP) defined by  projecting to the zero-momentum components, as $ C_{ij}(t-t') = \langle \bar O_i(t,\mathbf p = 0) O_j(t',\mathbf p = 0) \rangle$. The eigenvalues, $\lambda_n(t,t_0)$, and eigenvectors, $v_n(t,t_0)$, of $C(t)$ are defined to obey the relations
\begin{align}\label{eq:GEVP}
    C(t) v_n(t,t_0) = \lambda_n(t,t_0) C(t_0) v_n(t,t_0).
\end{align}
At large Euclidean times, the $n^{\rm th}$ eigenvalue relates to the $n^{\rm th}$ energy level in this channel, $E_n$, as
\begin{align}\label{eq:eigenvalue_asymptotics}
    \lambda_n(t \to \infty,t_0) = A_0 e^{-E_n (t-t_0)}\,.
\end{align}
We fit the eigenvalues to a lattice-periodic fit function. The fit-interval is chosen by a visual examination of the effective mass of the corresponding eigenvalue $\lambda_n(t,t_0)$, with $t_0$ fixed.

The determination of the mixing angle is based on Eq.~\eqref{eq:correlation_matrix},
restricted to unsmeared  operators~\cite{Bennett:2024wda}. The eigenvectors correspond to the matrix elements~\cite{Christ:2010dd}:
\begin{align} \label{eq:PS_matrix_elements}
    \begin{pmatrix}
        \langle 0  | O_{\eta^{\rm f }} | \eta'_l \rangle & \langle 0  | O_{\eta^{\rm as}} | \eta'_l \rangle    \\
        \langle 0  | O_{\eta^{\rm f}}  | \eta'_h \rangle & \langle 0  | O_{\eta^{\rm as}} | \eta'_h \rangle 
    \end{pmatrix}
    =
    \begin{pmatrix}
         A_{\rm f}^{\eta'_l} & A_{\rm as}^{\eta'_l}      \\
         A_{\rm f}^{\eta'_h} & A_{\rm as}^{\eta'_h}
    \end{pmatrix}
    \equiv
        \begin{pmatrix}
           A_{\eta'_l} \cos{\phi_{\eta'_l}} & A_{\eta'_l} \sin{\phi_{\eta'_l}} \\
         - A_{\eta'_h} \sin{\phi_{\eta'_h}} & A_{\eta'_h} \cos{\phi_{\eta'_h}}
    \end{pmatrix}\,.
\end{align}
In general, two mixing angles are needed to parameterize the matrix elements~\cite{Feldmann:1998vh}. We test whether the mixing could be described by a single mixing angle, $\phi$, by examining the following quantities
\begin{align}\label{eq:mixing_angles}
      \tan \phi_{\eta'_l} &\equiv \frac{A_{\rm as}^{\eta'_l}}{A_{\rm f}^{\eta'_l}}, & 
    - \tan \phi_{\eta'_h} &\equiv \frac{A_{\rm f}^{\eta'_h}}{A_{\rm as}^{\eta'_h}}, &
    - \tan^2 \phi         \equiv \frac{A_{\rm f}^{\eta'_h}A_{\rm as}^{\eta'_l}}{A_{\rm as}^{\eta'_h}A_{\rm f}^{\eta'_l}}.
\end{align}
If the system is parameterized by a single mixing angle, we should find $\phi_{\eta'_l} \approx \phi_{\eta'_h} \approx \phi$~\cite{Blossier:2009kd}.
For this study, we ignore other possible contributions, for instance due to mixing with pseudoscalar glueballs, or other excited states. 

\section{Numerical Setup}
\label{Sec:num}

Full details of the numerical simulation are provided in Refs.~\cite{Bennett:2024cqv,Bennett:2024wda}. We generate the gauge configurations on GPU-based machines, using the Grid software environment~\cite{Boyle:2015tjk,Boyle:2016lbp,Yamaguchi:2022feu}, extended to implement symplectic gauge theories~\cite{Bennett:2023gbe}. 
We generate gauge configurations on a lattice volume $L^3 \times T = a^4 N_s^3 \times N_t$ using the Wilson plaquette action and the standard Wilson discretization for the fermions without the clover term. We restrict ourselves to a single lattice spacing $(\beta = 6.5)$ and keep the bare mass of the three antisymmetric fermions fixed at $am_0^{\rm as}=-1.01$. We consider three different values of the bare fundamental fermion mass, $am_0^{\rm f}=-0.70, \,-0.71, \,-0.72$. For $am_0^{\rm f}=-0.71$, we study three different temporal extensions $N_t=48,\,64,\,90$. 

We  measure  correlation functions using the HiRep code~\cite{DelDebbio:2008zf,HiRepSUN,HiRepSpN}, on CPU-based machines. We implement both APE~\cite{APE:1987ehd}  and Wuppertal~\cite{Gusken:1989qx} smearings. APE smearing is performed with  smearing parameters $N_{\rm APE}=50$ and $\epsilon_{\rm APE}=0.4$. We use three different levels of Wuppertal smearing for each fermion representation, characterized by $N^{\rm smear}=0,\,40,\,80$ smearing steps, respectively, with Wuppertal smearing parameters $\epsilon_f=0.2$ and $\epsilon_{as}=0.12$ for the fundamental and antisymmetric fermions.
For the disconnected diagrams, we use $n_{\rm src}=64$ stochastic sources. 

The masses of all accessible flavour non-singlet states have been reported in Ref.~\cite{Bennett:2024cqv}. We denote the masses of the pseudoscalar mesons made of fundamental and antisymmetric fermions as $m_{\rm PS}$ and $m_{\rm ps}$, respectively. Similarly, for vector mesons, $m_{\rm V}$ and $m_{\rm v}$ stand for the flavoured, vector mesons.
On a finite lattice, an additive constant contribution appears in the correlation function for the pseudoscalars~\cite{Aoki:2007ka}, that we subtract by performing a numerical derivative at the level of the correlation matrix:
\begin{align}
        C_{ij}(t) \to \tilde C_{ij}(t) = \frac{C_{ij}(t-1) - C_{ij}(t+1)}{2}\,.
\end{align}
All statistical uncertainties are determined using the jackknife method.

\begin{figure}
    \centering
    \includegraphics[width=.4\textwidth]{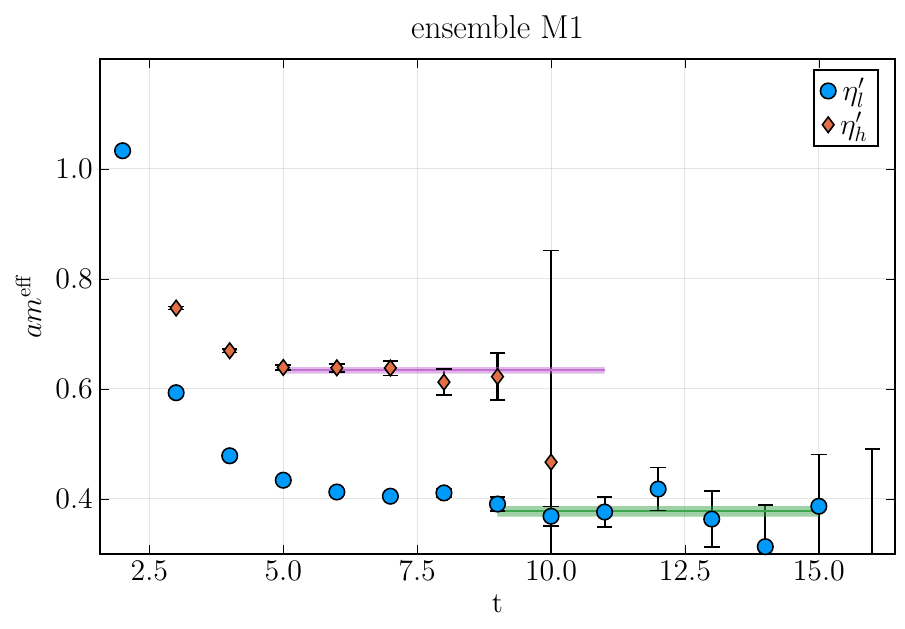} \includegraphics[width=.4\textwidth]{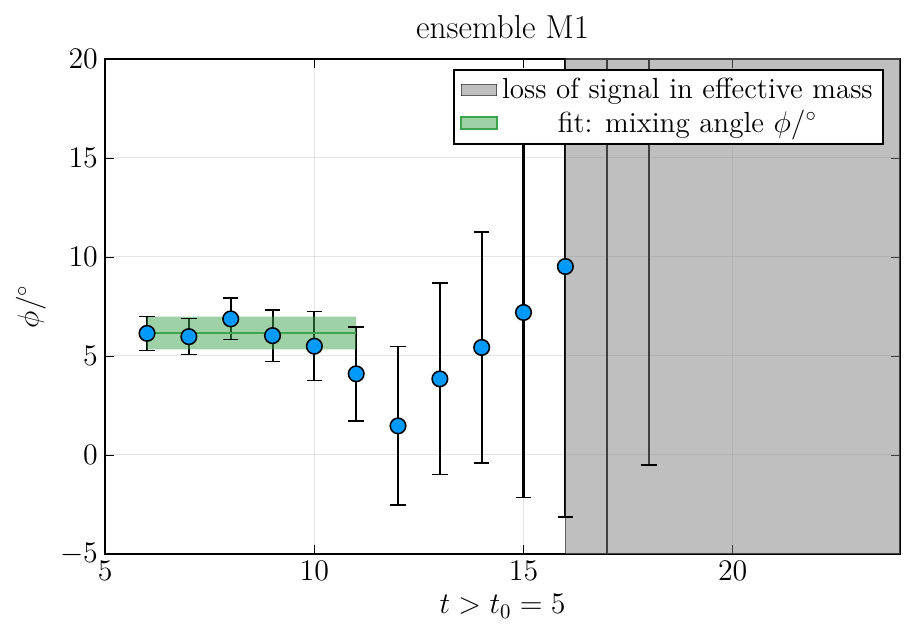} \\
    \includegraphics[width=.4\textwidth]{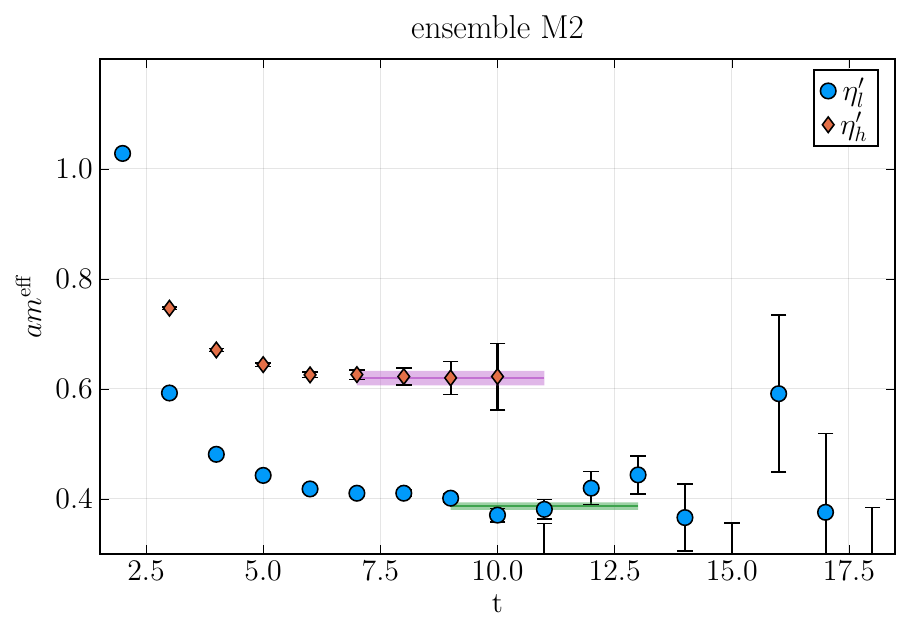} \includegraphics[width=.4\textwidth]{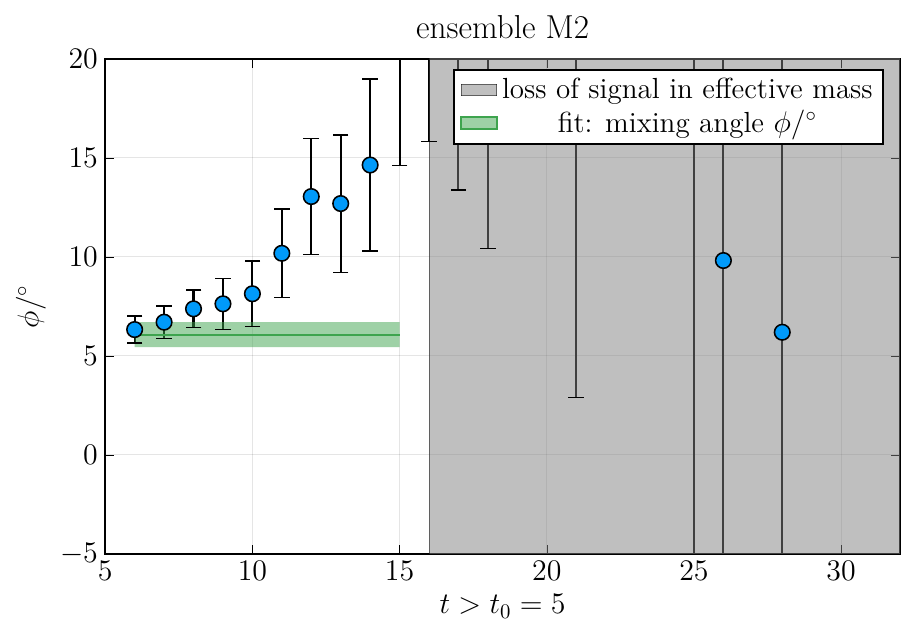} \\
    \includegraphics[width=.4\textwidth]{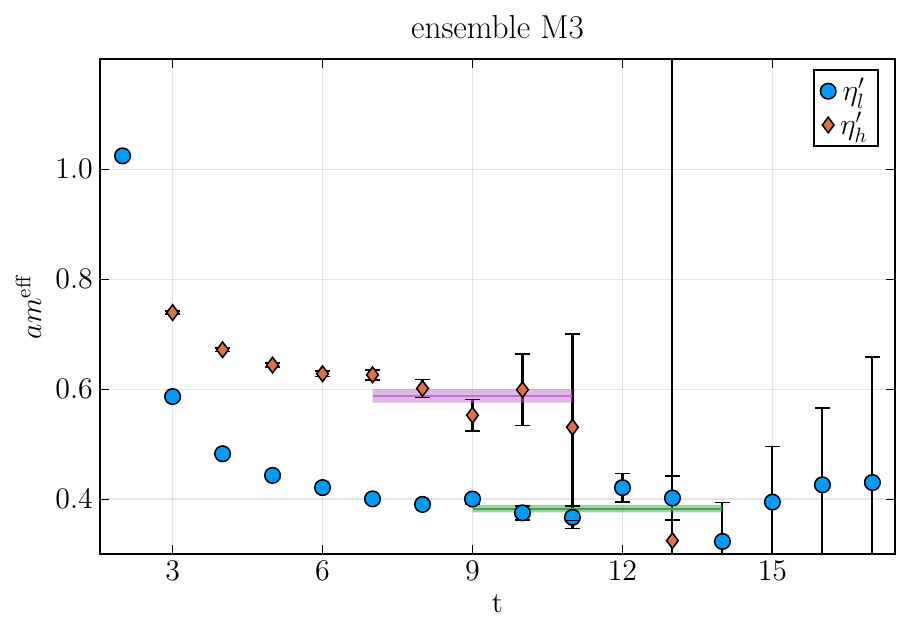} \includegraphics[width=.4\textwidth]{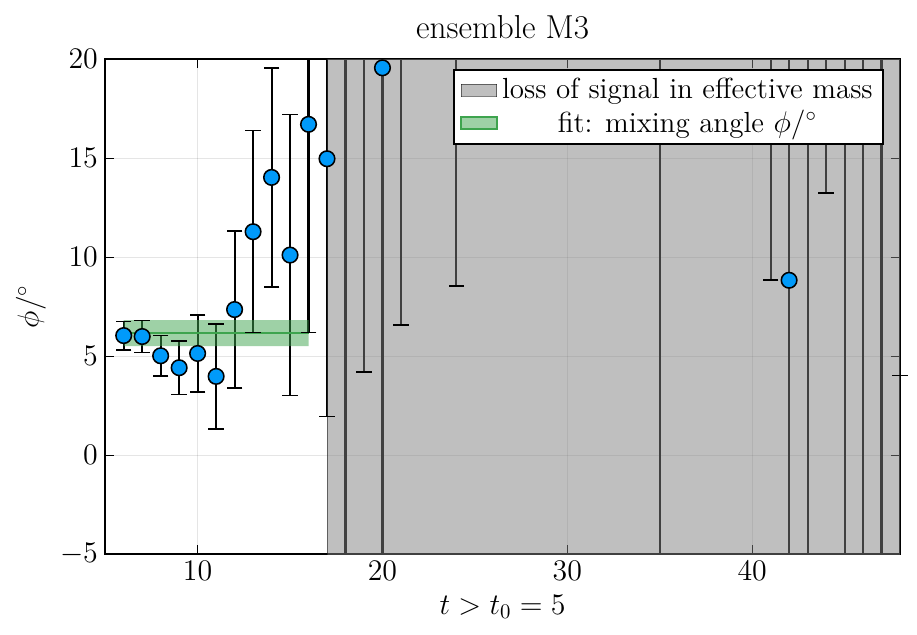} \\
    \includegraphics[width=.4\textwidth]{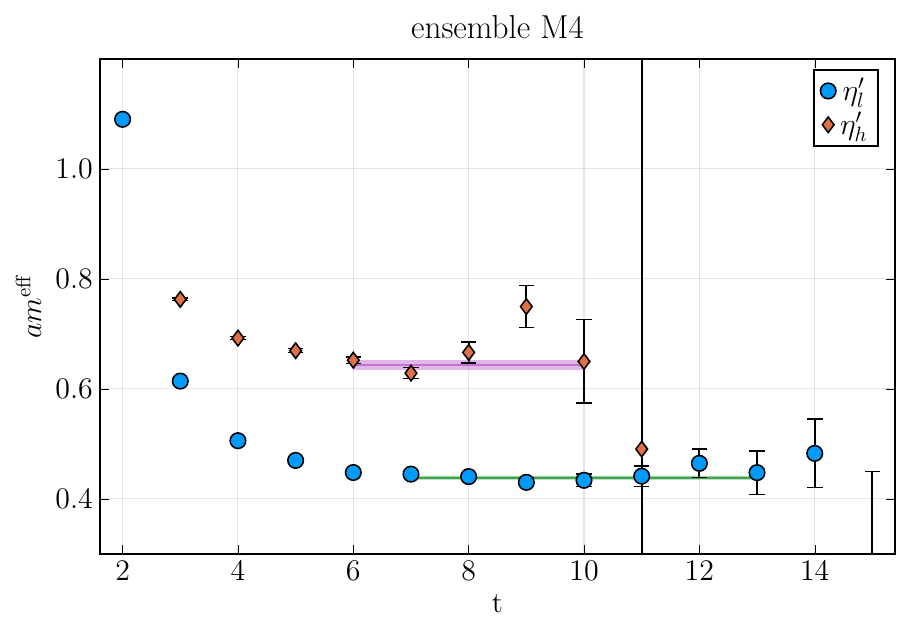} \includegraphics[width=.4\textwidth]{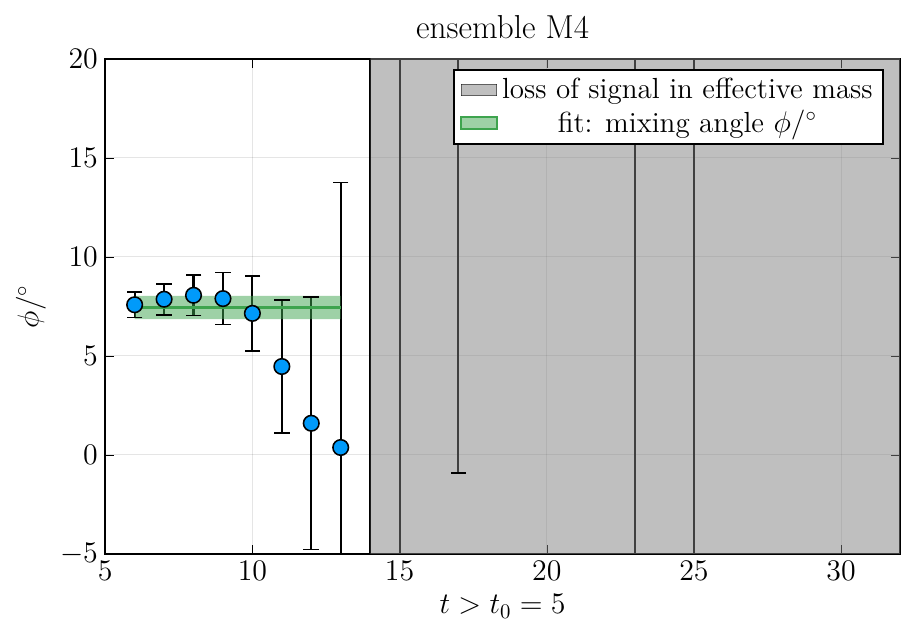} \\
    \includegraphics[width=.4\textwidth]{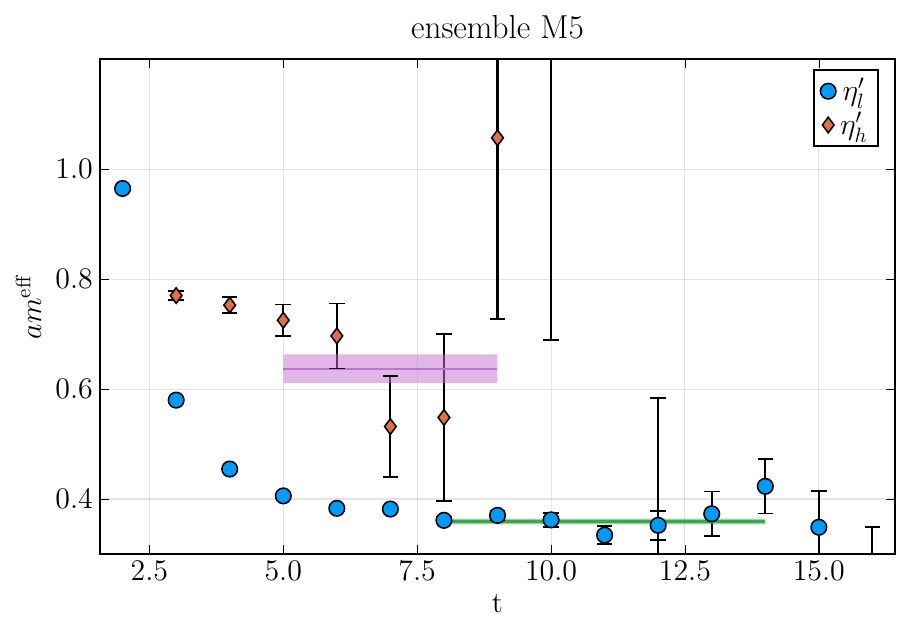} \includegraphics[width=.4\textwidth]{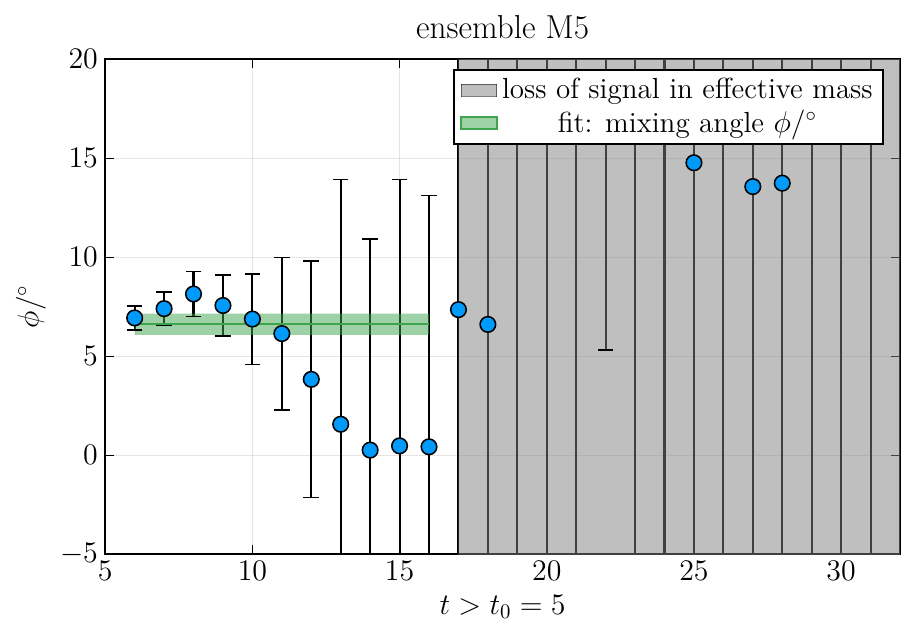} \\
    \caption{Effective mass (left) and effective average mixing angle (right) for each of the five ensembles.}
    \label{fig:effective_masses}
\end{figure}

\begin{table}
    \caption{For each available ensemble, lattice parameters and masses extracted for the pseudoscalar singlet states ($\eta'_l$ and $\eta'_h$) as well as the pseudoscalar non-singlets ($\rm PS$ and $\rm ps$) and vector non-singlets ($\rm V$ and $\rm v$).}
    \label{tab:masses}
    \begin{adjustbox}{width=\textwidth}
        \begin{tabular}{|c|c|c|c|c|c|c|c|c|c|c|c|}
	\hline
	Label&$\beta$&$N_t$&$N_l$&$am_0^{\rm f}$&$am_0^{\rm as}$&$am_{\eta^{\prime}_l}$&$am_{\eta^{\prime}_h}$&$am_{\rm PS}$&$am_{\rm ps}$&$am_{\rm V}$&$am_{\rm v}$\\
	\hline\hline
	M1&6.5&48&20&-0.71&-1.01&0.3769(96)&0.6334(59)&0.3639(14)&0.6001(11)&0.4030(33)&0.6452(18)\\
	M2&6.5&64&20&-0.71&-1.01&0.3867(68)&0.619(13)&0.3648(13)&0.59856(82)&0.4038(17)&0.6421(15)\\
	M3&6.5&96&20&-0.71&-1.01&0.3826(67)&0.588(12)&0.3652(16)&0.59940(79)&0.4040(18)&0.6467(21)\\
	M4&6.5&64&20&-0.7&-1.01&0.4381(33)&0.6433(88)&0.4067(13)&0.62426(85)&0.4476(17)&0.6742(13)\\
	M5&6.5&64&32&-0.72&-1.01&0.3591(53)&0.637(26)&0.31076(68)&0.57718(85)&0.3518(12)&0.6223(15)\\
	\hline\hline
\end{tabular}
    \end{adjustbox}

\end{table}
\section{Results}

We find  plateaux with modest time extent,  both in the ground state and the first excited state in the pseudoscalar flavour-singlet sector. The signal is marginally better for the ground state.  As a consequence, we expect the first excited state to be affected by systematic uncertainties due to the short plateaux. We perform fits over a minimum of four time slices.
In Fig.~\ref{fig:effective_masses}, we show the effective masses and the effective mixing angle for each of the five ensembles. The extracted energies are tabulated in Tab.~\ref{tab:masses}. We include the lightest non-singlet masses as a reference value. We further report all mixing angles, extracted according to Eq.~\eqref{eq:mixing_angles}, in Tab.~\ref{tab:mixing_results}.

\begin{table}
    \centering
    \begin{tabular}{|c|c|c|c|c|c|c|}
	\hline
    Label&$~~~~\beta~~~~$&$~~~~N_t~~~~$&$~~~~N_s~~~~$&$~~~~\phi/{}^{\circ}~~~~$&$~~~~\phi_{\eta'_l}/{}^{\circ}~~~~$&$~~~~\phi_{\eta'_h}/{}^{\circ}~~~~$ \\ \hline \hline 
	M1&6.5&48&20&6.15(83)&3.83(57)&9.8(1.1)\\
	M2&6.5&64&20&6.07(63)&3.74(43)&9.78(89)\\
	M3&6.5&96&20&6.16(66)&3.76(44)&10.00(92)\\
	M4&6.5&64&20&7.44(58)&4.77(42)&12.26(86)\\
	M5&6.5&64&32&6.61(54)&5.87(52)&7.67(64)\\
	\hline\hline
\end{tabular}
    \caption{Mixing angles according to Eq.~(\ref{eq:mixing_angles}), for each available ensemble } 
    \label{tab:mixing_results}
\end{table}

We find that the masses of the $\eta'_l$ and $\eta'_h$ are close to those of the pseudoscalar and vector mesons in the corresponding representations. This observation already suggests that these states are dominated by a single fermion representation. This hint is supported by the small mixing angles reported in Tab.~\ref{tab:mixing_results}. We also observe a large value of the ratio $m_{\rm PS}/m_{\rm V}$ close to unity, suggesting that the parameter space explored in this study corresponds to comparatively large values of the mass of the underlying hyperquarks (fermions). Indeed, we expect mixing effects to be dominated by the disconnected-diagram contributions to the off-diagonal terms of Eq.~\eqref{eq:correlation_matrix}, which are suppressed in the presence of  heavy fermions. 

\begin{figure}
    \centering
    \includegraphics[width=.48\textwidth]{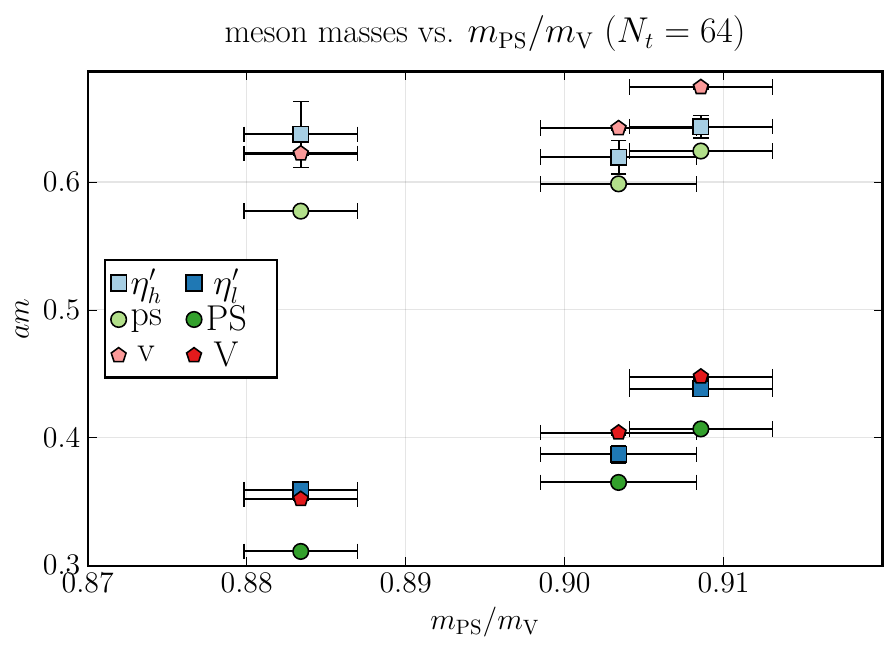}
    \includegraphics[width=.48\textwidth]{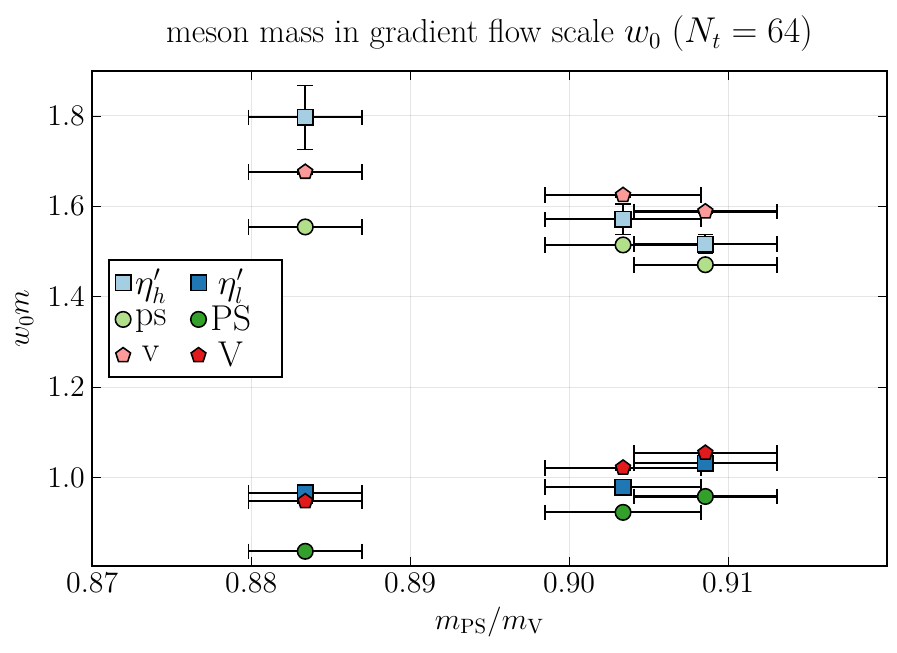}
    \caption{Mass spectrum of  light pseudoscalar and vector mesons. In the left panel we report the masses in lattice units, $a m$. In the right panel, the masses are expressed in units of the gradient flow scale, $m w_0$.}
    \label{fig:spectrum}
\end{figure}

The results in Tab.~\ref{tab:mixing_results} seem to suggest that  the parametrization of Eq.~\eqref{eq:PS_matrix_elements} by a single mixing angle be insufficient. While the overall angle is small throughout all ensembles, the difference between mixing angles $\phi_{\eta'_l}$ and $\phi_{\eta'_h}$ is statistically significant. We empirically found  that using a comparatively  large value of $t_0$ in the GEVP method  was advantageous in the determination of the mixing angles. We have chosen $t_0=5$ for this investigation. 

We observe, in Fig.~\ref{fig:spectrum}, that the masses of the mesons, expressed in lattice units, show some decrease when decreasing the mass of the fundamental fermions, even though only a modest change in $m_{\rm PS}/m_{\rm V}$ is observed. When plotting the fermion masses in units of the gradient flow scale, $w_0$, we observe that the masses of the $\eta_l$, $\rm PS$, and $\rm V$ states decrease as a function of $m_{\rm PS}/m_{\rm V}$, whereas the masses of the $\eta_h$, $\rm ps$, and $\rm v$ are increasing. A more extended discussion, including our operational choices for the definition of the Wilson flow scale, $w_0$, can be found in Refs.~\cite{Bennett:2024cqv,Bennett:2024wda}.

\section{Summary and outlook}

We performed the first direct determination of flavour-singlet meson masses and their mixing angles in a gauge theory with multiple fermion representations. The fermion masses are comparatively large, and as a consequence the mixing angle is small. The two singlet meson masses are  close to the pseudoscalar and vector flavoured meson masses, respectively.   The composition of the mass eigenstates is dominated by fermions of a single species.

Systematic uncertainties are probably sizeable, arising from  noisy signal, short plateaux and  coarse lattices available. Yet, this investigation is a stepping stone towards further, more sophisticated measurements. These include exploring lighter fermion masses, while combining  the scalar flavour-singlet meson channel with the glueball states. 

In order to map out the physics of composite Higgs models, an obvious next step is to decrease both the fundamental and antisymmetric fermion masses. Furthermore, the system should be studied at different lattice spacings. This could be either achieved by increasing the inverse gauge coupling, $\beta$, or by switching to a fermion action which has $\mathcal O(a)$ improvement.

\acknowledgments
EB and BL are supported by the EPSRC ExCALIBUR programme ExaTEPP (project EP/ X017168/1).  
EB, BL, MP, FZ are supported by the STFC Consolidated Grant No.ST/X000648/1. 
EB is supported by the STFC Research Software Engineering Fellowship EP/V052489/1.
NF is supported by the STFC Consolidated Grant No. ST/X508834/1. 
DKH is supported by Basic Science Research Program through the National Research Foundation of Korea (NRF) funded by the Ministry of Education (NRF-2017R1D1A1B06033701) and the NRF grant MSIT 2021R1A4A5031460 funded by the Korean government.
JWL is supported by IBS under the project code IBS-R018-D1. 
HH and CJDL are supported by the  Taiwanese MoST grant 109-2112-M-009-006-MY3 and NSTC grant 112-2112-M-A49-021-MY3.
CJDL is also supported by Grants No. 112-2639-M-002-006-ASP and No. 113-2119-M-007-013. 
BL and MP have been supported by the STFC Consolidated Grant No. ST/T000813/1 and by the the European Research Council (ERC) under the European Union’s Horizon 2020 research and innovation program under Grant Agreement No. 813942. 
DV is supported by the STFC under Consolidated Grant No. ST/X000680/1.

Numerical simulations have been performed on the DiRAC Extreme Scaling service at The University of Edinburgh, and on the DiRAC Data Intensive service at Leicester. 
The DiRAC Extreme Scaling service is operated by the Edinburgh Parallel Computing Centre on behalf of the STFC DiRAC HPC Facility (www.dirac.ac.uk). 
This equipment was funded by BEIS capital funding via STFC capital grant ST/R00238X/1 and STFC DiRAC Operations grant ST/R001006/1.
DiRAC is part of the UKRI Digital Research Infrastructure

{\bf Research Data Access Statement}---The data and analysis code generated for these proceedings and the extended publication in Ref.~\cite{Bennett:2024wda} can be downloaded from  Ref.~\cite{bennett_2024_11370542} and~\cite{bennett_2024_12748631}, respectively.

{\bf Open Access Statement}---For the purpose of open access, the authors have applied a Creative Commons  Attribution (CC BY) licence  to any Author Accepted Manuscript version arising.

\newpage
\bibliographystyle{JHEP}
\bibliography{bibliography.bib}

\providecommand{\href}[2]{#2}\begingroup\raggedright\begin{thebibliography}{10}

\bibitem{Kaplan:1983fs}
D.B.~Kaplan and H.~Georgi, \emph{{SU(2) x U(1) Breaking by Vacuum Misalignment}}, \href{https://doi.org/10.1016/0370-2693(84)91177-8}{\emph{Phys. Lett. B} {\bfseries 136} (1984) 183}.

\bibitem{Georgi:1984af}
H.~Georgi and D.B.~Kaplan, \emph{{Composite Higgs and Custodial SU(2)}}, \href{https://doi.org/10.1016/0370-2693(84)90341-1}{\emph{Phys. Lett. B} {\bfseries 145} (1984) 216}.

\bibitem{Dugan:1984hq}
M.J.~Dugan, H.~Georgi and D.B.~Kaplan, \emph{{Anatomy of a Composite Higgs Model}}, \href{https://doi.org/10.1016/0550-3213(85)90221-4}{\emph{Nucl. Phys. B} {\bfseries 254} (1985) 299}.

\bibitem{Kaplan:1991dc}
D.B.~Kaplan, \emph{{Flavor at SSC energies: A New mechanism for dynamically generated fermion masses}}, \href{https://doi.org/10.1016/S0550-3213(05)80021-5}{\emph{Nucl. Phys. B} {\bfseries 365} (1991) 259}.

\bibitem{Vecchi:2015fma}
L.~Vecchi, \emph{{A dangerous irrelevant UV-completion of the composite Higgs}}, \href{https://doi.org/10.1007/JHEP02(2017)094}{\emph{JHEP} {\bfseries 02} (2017) 094} [\href{https://arxiv.org/abs/1506.00623}{{\ttfamily 1506.00623}}].

\bibitem{Barnard:2013zea}
J.~Barnard, T.~Gherghetta and T.S.~Ray, \emph{{UV descriptions of composite Higgs models without elementary scalars}}, \href{https://doi.org/10.1007/JHEP02(2014)002}{\emph{JHEP} {\bfseries 02} (2014) 002} [\href{https://arxiv.org/abs/1311.6562}{{\ttfamily 1311.6562}}].

\bibitem{Panico:2015jxa}
G.~Panico and A.~Wulzer, \emph{{The Composite Nambu-Goldstone Higgs}},  \href{https://arxiv.org/abs/1506.01961}{{\ttfamily 1506.01961}}.

\bibitem{Witzel:2019jbe}
O.~Witzel, \emph{{Review on Composite Higgs Models}}, \href{https://doi.org/10.22323/1.334.0006}{\emph{PoS} {\bfseries LATTICE2018} (2019) 006} [\href{https://arxiv.org/abs/1901.08216}{{\ttfamily 1901.08216}}].

\bibitem{Cacciapaglia:2020kgq}
G.~Cacciapaglia, C.~Pica and F.~Sannino, \emph{{Fundamental Composite Dynamics: A Review}}, \href{https://doi.org/10.1016/j.physrep.2020.07.002}{\emph{Phys. Rept.} {\bfseries 877} (2020) 1} [\href{https://arxiv.org/abs/2002.04914}{{\ttfamily 2002.04914}}].

\bibitem{Ferretti:2013kya}
G.~Ferretti and D.~Karateev, \emph{{Fermionic UV completions of Composite Higgs models}}, \href{https://doi.org/10.1007/JHEP03(2014)077}{\emph{JHEP} {\bfseries 03} (2014) 077} [\href{https://arxiv.org/abs/1312.5330}{{\ttfamily 1312.5330}}].

\bibitem{Ferretti:2016upr}
G.~Ferretti, \emph{{Gauge theories of Partial Compositeness: Scenarios for Run-II of the LHC}}, \href{https://doi.org/10.1007/JHEP06(2016)107}{\emph{JHEP} {\bfseries 06} (2016) 107} [\href{https://arxiv.org/abs/1604.06467}{{\ttfamily 1604.06467}}].

\bibitem{Cacciapaglia:2019bqz}
G.~Cacciapaglia, G.~Ferretti, T.~Flacke and H.~Ser\^odio, \emph{{Light scalars in composite Higgs models}}, \href{https://doi.org/10.3389/fphy.2019.00022}{\emph{Front. in Phys.} {\bfseries 7} (2019) 22} [\href{https://arxiv.org/abs/1902.06890}{{\ttfamily 1902.06890}}].

\bibitem{Bennett:2017kga}
E.~Bennett, D.K.~Hong, J.-W.~Lee, C.J.D.~Lin, B.~Lucini, M.~Piai et~al., \emph{{Sp(4) gauge theory on the lattice: towards SU(4)/Sp(4) composite Higgs (and beyond)}}, \href{https://doi.org/10.1007/JHEP03(2018)185}{\emph{JHEP} {\bfseries 03} (2018) 185} [\href{https://arxiv.org/abs/1712.04220}{{\ttfamily 1712.04220}}].

\bibitem{Bennett:2019jzz}
E.~Bennett, D.K.~Hong, J.-W.~Lee, C.J.D.~Lin, B.~Lucini, M.~Piai et~al., \emph{{Sp(4) gauge theories on the lattice: $N_f=2$ dynamical fundamental fermions}}, \href{https://doi.org/10.1007/JHEP12(2019)053}{\emph{JHEP} {\bfseries 12} (2019) 053} [\href{https://arxiv.org/abs/1909.12662}{{\ttfamily 1909.12662}}].

\bibitem{Bennett:2019cxd}
E.~Bennett, D.K.~Hong, J.-W.~Lee, C.-J.D.~Lin, B.~Lucini, M.~Mesiti et~al., \emph{{$Sp(4)$ gauge theories on the lattice: quenched fundamental and antisymmetric fermions}}, \href{https://doi.org/10.1103/PhysRevD.101.074516}{\emph{Phys. Rev. D} {\bfseries 101} (2020) 074516} [\href{https://arxiv.org/abs/1912.06505}{{\ttfamily 1912.06505}}].

\bibitem{Bennett:2020hqd}
E.~Bennett, J.~Holligan, D.K.~Hong, J.-W.~Lee, C.J.D.~Lin, B.~Lucini et~al., \emph{{Color dependence of tensor and scalar glueball masses in Yang-Mills theories}}, \href{https://doi.org/10.1103/PhysRevD.102.011501}{\emph{Phys. Rev. D} {\bfseries 102} (2020) 011501} [\href{https://arxiv.org/abs/2004.11063}{{\ttfamily 2004.11063}}].

\bibitem{Bennett:2020qtj}
E.~Bennett, J.~Holligan, D.K.~Hong, J.-W.~Lee, C.J.D.~Lin, B.~Lucini et~al., \emph{{Glueballs and strings in $Sp(2N)$ Yang-Mills theories}}, \href{https://doi.org/10.1103/PhysRevD.103.054509}{\emph{Phys. Rev. D} {\bfseries 103} (2021) 054509} [\href{https://arxiv.org/abs/2010.15781}{{\ttfamily 2010.15781}}].

\bibitem{Bennett:2022yfa}
E.~Bennett, D.K.~Hong, H.~Hsiao, J.-W.~Lee, C.J.D.~Lin, B.~Lucini et~al., \emph{{Lattice studies of the Sp(4) gauge theory with two fundamental and three antisymmetric Dirac fermions}}, \href{https://doi.org/10.1103/PhysRevD.106.014501}{\emph{Phys. Rev. D} {\bfseries 106} (2022) 014501} [\href{https://arxiv.org/abs/2202.05516}{{\ttfamily 2202.05516}}].

\bibitem{Bennett:2022gdz}
E.~Bennett, D.K.~Hong, J.-W.~Lee, C.J.D.~Lin, B.~Lucini, M.~Piai et~al., \emph{{Color dependence of the topological susceptibility in Yang-Mills theories}}, \href{https://doi.org/10.1016/j.physletb.2022.137504}{\emph{Phys. Lett. B} {\bfseries 835} (2022) 137504} [\href{https://arxiv.org/abs/2205.09254}{{\ttfamily 2205.09254}}].

\bibitem{Bennett:2022ftz}
E.~Bennett, D.K.~Hong, J.-W.~Lee, C.J.D.~Lin, B.~Lucini, M.~Piai et~al., \emph{{Sp(2N) Yang-Mills theories on the lattice: Scale setting and topology}}, \href{https://doi.org/10.1103/PhysRevD.106.094503}{\emph{Phys. Rev. D} {\bfseries 106} (2022) 094503} [\href{https://arxiv.org/abs/2205.09364}{{\ttfamily 2205.09364}}].

\bibitem{Bennett:2023wjw}
E.~Bennett, J.~Holligan, D.K.~Hong, H.~Hsiao, J.-W.~Lee, C.J.D.~Lin et~al., \emph{{Sp(2N) Lattice Gauge Theories and Extensions of the Standard Model of Particle Physics}}, \href{https://doi.org/10.3390/universe9050236}{\emph{Universe} {\bfseries 9} (2023) 236} [\href{https://arxiv.org/abs/2304.01070}{{\ttfamily 2304.01070}}].

\bibitem{Bennett:2023gbe}
E.~Bennett et~al., \emph{{Symplectic lattice gauge theories in the grid framework: Approaching the conformal window}}, \href{https://doi.org/10.1103/PhysRevD.108.094508}{\emph{Phys. Rev. D} {\bfseries 108} (2023) 094508} [\href{https://arxiv.org/abs/2306.11649}{{\ttfamily 2306.11649}}].

\bibitem{Bennett:2023mhh}
E.~Bennett, D.K.~Hong, H.~Hsiao, J.-W.~Lee, C.J.D.~Lin, B.~Lucini et~al., \emph{{Lattice investigations of the chimera baryon spectrum in the Sp(4) gauge theory}}, \href{https://doi.org/10.1103/PhysRevD.109.094512}{\emph{Phys. Rev. D} {\bfseries 109} (2024) 094512} [\href{https://arxiv.org/abs/2311.14663}{{\ttfamily 2311.14663}}].

\bibitem{Bennett:2023qwx}
E.~Bennett, J.~Holligan, D.K.~Hong, J.-W.~Lee, C.J.D.~Lin, B.~Lucini et~al., \emph{{Spectrum of mesons in quenched Sp(2N) gauge theories}}, \href{https://doi.org/10.1103/PhysRevD.109.094517}{\emph{Phys. Rev. D} {\bfseries 109} (2024) 094517} [\href{https://arxiv.org/abs/2312.08465}{{\ttfamily 2312.08465}}].

\bibitem{Bennett:2024cqv}
E.~Bennett et~al., \emph{{Meson spectroscopy from spectral densities in lattice gauge theories}},  \href{https://arxiv.org/abs/2405.01388}{{\ttfamily 2405.01388}}.

\bibitem{Bennett:2024wda}
E.~Bennett, N.~Forzano, D.K.~Hong, H.~Hsiao, J.-W.~Lee, C.J.D.~Lin et~al., \emph{{Mixing between flavor singlets in lattice gauge theories coupled to matter fields in multiple representations}}, \href{https://doi.org/10.1103/PhysRevD.110.074504}{\emph{Phys. Rev. D} {\bfseries 110} (2024) 074504} [\href{https://arxiv.org/abs/2405.05765}{{\ttfamily 2405.05765}}].

\bibitem{Kulkarni:2022bvh}
S.~Kulkarni, A.~Maas, S.~Mee, M.~Nikolic, J.~Pradler and F.~Zierler, \emph{{Low-energy effective description of dark $Sp(4)$ theories}}, \href{https://doi.org/10.21468/SciPostPhys.14.3.044}{\emph{SciPost Phys.} {\bfseries 14} (2023) 044} [\href{https://arxiv.org/abs/2202.05191}{{\ttfamily 2202.05191}}].

\bibitem{Bennett:2023rsl}
E.~Bennett, H.~Hsiao, J.-W.~Lee, B.~Lucini, A.~Maas, M.~Piai et~al., \emph{{Singlets in gauge theories with fundamental matter}}, {\emph{pre-print} (2023) } [\href{https://arxiv.org/abs/2304.07191}{{\ttfamily 2304.07191}}].

\bibitem{Dengler:2024maq}
Y.~Dengler, A.~Maas and F.~Zierler, \emph{{Scattering of dark pions in Sp(4) gauge theory}}, \href{https://doi.org/10.1103/PhysRevD.110.054513}{\emph{Phys. Rev. D} {\bfseries 110} (2024) 054513} [\href{https://arxiv.org/abs/2405.06506}{{\ttfamily 2405.06506}}].

\bibitem{Bennett:2024bhy}
E.~Bennett, B.~Lucini, D.~Mason, M.~Piai, E.~Rinaldi and D.~Vadacchino, \emph{{The density of states method for symplectic gauge theories at finite temperature}},  \href{https://arxiv.org/abs/2409.19426}{{\ttfamily 2409.19426}}.

\bibitem{DeGrand:2016mxr}
T.A.~DeGrand, D.~Hackett, W.I.~Jay, E.T.~Neil, Y.~Shamir and B.~Svetitsky, \emph{{Towards Partial Compositeness on the Lattice: Baryons with Fermions in Multiple Representations}}, \href{https://doi.org/10.22323/1.256.0219}{\emph{PoS} {\bfseries LATTICE2016} (2016) 219} [\href{https://arxiv.org/abs/1610.06465}{{\ttfamily 1610.06465}}].

\bibitem{Ayyar:2017qdf}
V.~Ayyar, T.~DeGrand, M.~Golterman, D.C.~Hackett, W.I.~Jay, E.T.~Neil et~al., \emph{{Spectroscopy of SU(4) composite Higgs theory with two distinct fermion representations}}, \href{https://doi.org/10.1103/PhysRevD.97.074505}{\emph{Phys. Rev. D} {\bfseries 97} (2018) 074505} [\href{https://arxiv.org/abs/1710.00806}{{\ttfamily 1710.00806}}].

\bibitem{Ayyar:2017uqh}
V.~Ayyar, T.~DeGrand, D.C.~Hackett, W.I.~Jay, E.T.~Neil, Y.~Shamir et~al., \emph{{Chiral Transition of SU(4) Gauge Theory with Fermions in Multiple Representations}}, \href{https://doi.org/10.1051/epjconf/201817508026}{\emph{EPJ Web Conf.} {\bfseries 175} (2018) 08026} [\href{https://arxiv.org/abs/1709.06190}{{\ttfamily 1709.06190}}].

\bibitem{Ayyar:2017vsu}
V.~Ayyar, D.~Hackett, W.~Jay and E.~Neil, \emph{{Confinement study of an SU(4) gauge theory with fermions in multiple representations}}, \href{https://doi.org/10.1051/epjconf/201817508025}{\emph{EPJ Web Conf.} {\bfseries 175} (2018) 08025} [\href{https://arxiv.org/abs/1710.03257}{{\ttfamily 1710.03257}}].

\bibitem{Ayyar:2018ppa}
V.~Ayyar, T.~DeGrand, D.C.~Hackett, W.I.~Jay, E.T.~Neil, Y.~Shamir et~al., \emph{{Finite-temperature phase structure of SU(4) gauge theory with multiple fermion representations}}, \href{https://doi.org/10.1103/PhysRevD.97.114502}{\emph{Phys. Rev. D} {\bfseries 97} (2018) 114502} [\href{https://arxiv.org/abs/1802.09644}{{\ttfamily 1802.09644}}].

\bibitem{Ayyar:2018zuk}
V.~Ayyar, T.~Degrand, D.C.~Hackett, W.I.~Jay, E.T.~Neil, Y.~Shamir et~al., \emph{{Baryon spectrum of SU(4) composite Higgs theory with two distinct fermion representations}}, \href{https://doi.org/10.1103/PhysRevD.97.114505}{\emph{Phys. Rev. D} {\bfseries 97} (2018) 114505} [\href{https://arxiv.org/abs/1801.05809}{{\ttfamily 1801.05809}}].

\bibitem{Cossu:2019hse}
G.~Cossu, L.~Del~Debbio, M.~Panero and D.~Preti, \emph{{Strong dynamics with matter in multiple representations: $\mathrm {SU}(4)$ gauge theory with fundamental and sextet fermions}}, \href{https://doi.org/10.1140/epjc/s10052-019-7137-1}{\emph{Eur. Phys. J. C} {\bfseries 79} (2019) 638} [\href{https://arxiv.org/abs/1904.08885}{{\ttfamily 1904.08885}}].

\bibitem{Lupo:2021nzv}
A.~Lupo, M.~Panero, N.~Tantalo and L.~Del~Debbio, \emph{{Spectral reconstruction in SU(4) gauge theory with fermions in multiple representations}}, \href{https://doi.org/10.22323/1.396.0092}{\emph{PoS} {\bfseries LATTICE2021} (2022) 092} [\href{https://arxiv.org/abs/2112.01158}{{\ttfamily 2112.01158}}].

\bibitem{DelDebbio:2022qgu}
L.~Del~Debbio, A.~Lupo, M.~Panero and N.~Tantalo, \emph{{Multi-representation dynamics of SU(4) composite Higgs models: chiral limit and spectral reconstructions}}, \href{https://doi.org/10.1140/epjc/s10052-023-11363-8}{\emph{Eur. Phys. J. C} {\bfseries 83} (2023) 220} [\href{https://arxiv.org/abs/2211.09581}{{\ttfamily 2211.09581}}].

\bibitem{Golterman:2020pyx}
M.~Golterman, W.I.~Jay, E.T.~Neil, Y.~Shamir and B.~Svetitsky, \emph{{Low-energy constant $L_{10}$ in a two-representation lattice theory}}, \href{https://doi.org/10.1103/PhysRevD.103.074509}{\emph{Phys. Rev. D} {\bfseries 103} (2021) 074509} [\href{https://arxiv.org/abs/2010.01920}{{\ttfamily 2010.01920}}].

\bibitem{Hasenfratz:2023sqa}
A.~Hasenfratz, E.T.~Neil, Y.~Shamir, B.~Svetitsky and O.~Witzel, \emph{{Infrared fixed point and anomalous dimensions in a composite Higgs model}}, \href{https://doi.org/10.1103/PhysRevD.107.114504}{\emph{Phys. Rev. D} {\bfseries 107} (2023) 114504} [\href{https://arxiv.org/abs/2304.11729}{{\ttfamily 2304.11729}}].

\bibitem{Bergner:2020mwl}
G.~Bergner and S.~Piemonte, \emph{{Lattice simulations of a gauge theory with mixed adjoint-fundamental matter}}, \href{https://doi.org/10.1103/PhysRevD.103.014503}{\emph{Phys. Rev. D} {\bfseries 103} (2021) 014503} [\href{https://arxiv.org/abs/2008.02855}{{\ttfamily 2008.02855}}].

\bibitem{Bergner:2021ivi}
G.~Bergner and S.~Piemonte, \emph{{Mixed adjoint-fundamental matter and applications towards SQCD and beyond}}, \href{https://doi.org/10.22323/1.396.0242}{\emph{PoS} {\bfseries LATTICE2021} (2022) 242} [\href{https://arxiv.org/abs/2111.15335}{{\ttfamily 2111.15335}}].

\bibitem{Belyaev:2015hgo}
A.~Belyaev, G.~Cacciapaglia, H.~Cai, T.~Flacke, A.~Parolini and H.~Ser\^odio, \emph{{Singlets in composite Higgs models in light of the LHC 750 GeV diphoton excess}}, \href{https://doi.org/10.1103/PhysRevD.94.015004}{\emph{Phys. Rev. D} {\bfseries 94} (2016) 015004} [\href{https://arxiv.org/abs/1512.07242}{{\ttfamily 1512.07242}}].

\bibitem{DeGrand:2016pgq}
T.~DeGrand, M.~Golterman, E.T.~Neil and Y.~Shamir, \emph{{One-loop Chiral Perturbation Theory with two fermion representations}}, \href{https://doi.org/10.1103/PhysRevD.94.025020}{\emph{Phys. Rev. D} {\bfseries 94} (2016) 025020} [\href{https://arxiv.org/abs/1605.07738}{{\ttfamily 1605.07738}}].

\bibitem{Witten:1979vv}
E.~Witten, \emph{{Current Algebra Theorems for the U(1) Goldstone Boson}}, \href{https://doi.org/10.1016/0550-3213(79)90031-2}{\emph{Nucl. Phys. B} {\bfseries 156} (1979) 269}.

\bibitem{Witten:1978bc}
E.~Witten, \emph{{Instantons, the Quark Model, and the 1/n Expansion}}, \href{https://doi.org/10.1016/0550-3213(79)90243-8}{\emph{Nucl. Phys. B} {\bfseries 149} (1979) 285}.

\bibitem{Veneziano:1979ec}
G.~Veneziano, \emph{{U(1) Without Instantons}}, \href{https://doi.org/10.1016/0550-3213(79)90332-8}{\emph{Nucl. Phys. B} {\bfseries 159} (1979) 213}.

\bibitem{BuarqueFranzosi:2021kky}
D.~Buarque~Franzosi, G.~Cacciapaglia, X.~Cid~Vidal, G.~Ferretti, T.~Flacke and C.~V\'azquez~Sierra, \emph{{Exploring new possibilities to discover a light pseudo-scalar at LHCb}}, \href{https://doi.org/10.1140/epjc/s10052-021-09930-y}{\emph{Eur. Phys. J. C} {\bfseries 82} (2022) 3} [\href{https://arxiv.org/abs/2106.12615}{{\ttfamily 2106.12615}}].

\bibitem{Belyaev:2016ftv}
A.~Belyaev, G.~Cacciapaglia, H.~Cai, G.~Ferretti, T.~Flacke, A.~Parolini et~al., \emph{{Di-boson signatures as Standard Candles for Partial Compositeness}}, \href{https://doi.org/10.1007/JHEP01(2017)094}{\emph{JHEP} {\bfseries 01} (2017) 094} [\href{https://arxiv.org/abs/1610.06591}{{\ttfamily 1610.06591}}].

\bibitem{Christ:2010dd}
N.H.~Christ, C.~Dawson, T.~Izubuchi, C.~Jung, Q.~Liu, R.D.~Mawhinney et~al., \emph{{The $\eta$ and $\eta^\prime$ mesons from Lattice QCD}}, \href{https://doi.org/10.1103/PhysRevLett.105.241601}{\emph{Phys. Rev. Lett.} {\bfseries 105} (2010) 241601} [\href{https://arxiv.org/abs/1002.2999}{{\ttfamily 1002.2999}}].

\bibitem{Feldmann:1998vh}
T.~Feldmann, P.~Kroll and B.~Stech, \emph{{Mixing and decay constants of pseudoscalar mesons}}, \href{https://doi.org/10.1103/PhysRevD.58.114006}{\emph{Phys. Rev. D} {\bfseries 58} (1998) 114006} [\href{https://arxiv.org/abs/hep-ph/9802409}{{\ttfamily hep-ph/9802409}}].

\bibitem{Blossier:2009kd}
B.~Blossier, M.~Della~Morte, G.~von Hippel, T.~Mendes and R.~Sommer, \emph{{On the generalized eigenvalue method for energies and matrix elements in lattice field theory}}, \href{https://doi.org/10.1088/1126-6708/2009/04/094}{\emph{JHEP} {\bfseries 04} (2009) 094} [\href{https://arxiv.org/abs/0902.1265}{{\ttfamily 0902.1265}}].

\bibitem{Boyle:2015tjk}
P.~Boyle, A.~Yamaguchi, G.~Cossu and A.~Portelli, \emph{{Grid: A next generation data parallel C++ QCD library}},  12, 2015.

\bibitem{Boyle:2016lbp}
P.A.~Boyle, G.~Cossu, A.~Yamaguchi and A.~Portelli, \emph{{Grid: A next generation data parallel C++ QCD library}}, \href{https://doi.org/10.22323/1.251.0023}{\emph{PoS} {\bfseries LATTICE2015} (2016) 023}.

\bibitem{Yamaguchi:2022feu}
A.~Yamaguchi, P.~Boyle, G.~Cossu, G.~Filaci, C.~Lehner and A.~Portelli, \emph{{Grid: OneCode and FourAPIs}}, \href{https://doi.org/10.22323/1.396.0035}{\emph{PoS} {\bfseries LATTICE2021} (2022) 035} [\href{https://arxiv.org/abs/2203.06777}{{\ttfamily 2203.06777}}].

\bibitem{DelDebbio:2008zf}
L.~Del~Debbio, A.~Patella and C.~Pica, \emph{{Higher representations on the lattice: Numerical simulations. SU(2) with adjoint fermions}}, \href{https://doi.org/10.1103/PhysRevD.81.094503}{\emph{Phys. Rev. D} {\bfseries 81} (2010) 094503} [\href{https://arxiv.org/abs/0805.2058}{{\ttfamily 0805.2058}}].

\bibitem{HiRepSUN}
``{G}it{H}ub - claudiopica/{H}i{R}ep: {H}i{R}ep repository --- github.com.'' \url{https://github.com/claudiopica/HiRep}, accessed September 2024.

\bibitem{HiRepSpN}
``{G}it{H}ub - sa2c/{H}i{R}ep: {H}i{R}ep repository --- github.com.'' \url{https://github.com/sa2c/HiRep}, accessed September 2024.

\bibitem{APE:1987ehd}
{\scshape APE} collaboration, \emph{{Glueball Masses and String Tension in Lattice QCD}}, \href{https://doi.org/10.1016/0370-2693(87)91160-9}{\emph{Phys. Lett. B} {\bfseries 192} (1987) 163}.

\bibitem{Gusken:1989qx}
S.~Gusken, \emph{{A Study of smearing techniques for hadron correlation functions}}, \href{https://doi.org/10.1016/0920-5632(90)90273-W}{\emph{Nucl. Phys. B Proc. Suppl.} {\bfseries 17} (1990) 361}.

\bibitem{Aoki:2007ka}
S.~Aoki, H.~Fukaya, S.~Hashimoto and T.~Onogi, \emph{{Finite volume QCD at fixed topological charge}}, \href{https://doi.org/10.1103/PhysRevD.76.054508}{\emph{Phys. Rev. D} {\bfseries 76} (2007) 054508} [\href{https://arxiv.org/abs/0707.0396}{{\ttfamily 0707.0396}}].

\bibitem{bennett_2024_11370542}
E.~Bennett, N.~Forzano, D.K.~Hong, H.~Hsiao, J.-W.~Lee, C.-J.D.~Lin et~al., \emph{{On the mixing between flavor singlets in lattice gauge theories coupled to matter fields in multiple representations - data release}},  July, 2024.
\newblock 10.5281/zenodo.11370542.

\bibitem{bennett_2024_12748631}
E.~Bennett, N.~Forzano, D.K.~Hong, H.~Hsiao, J.-W.~Lee, C.-J.D.~Lin et~al., \emph{{On the mixing between flavor singlets in lattice gauge theories coupled to matter fields in multiple representations - workflow release}},  July, 2024.
\newblock 10.5281/zenodo.12748631.

\end{thebibliography}\endgroup

\end{document}